# Photoelectronic scoping of adatoms, atomic vacancies, and the outermost layer of a surface

Yan Wang,[1] Yanguang Nie,[2] Jisheng Pan,[3] Wei Qin,[4] Zhaofeng Zhou,[4] Weitao Zheng,[5] Chang Q Sun,[2*]

[1] School of Information and Electronic Engineering, Hunan University of Science and Technology, Xiangtan 411201, China; [2]School of Electrical and Electronic Engineering, Nanyang Technological University, Singapore 639798; [3]Institute of Materials Research and Engineering, A*Star, Singapore 117602; [4]Faculty of Materials, Photoelectronics and Physics, Xiangtan University, Changsha 400073, China; [5]School of Materials Science, Jilin University, Changchun 130012, China

An effective yet simple means disclosed herewith has allowed us to gain the atomistic, local, and quantitative information of bonds and electrons at sites surrounding undercoordinated atoms, complementing the scanning tunneling microscopy/spectroscopy and photoelectron spectroscopy (XPS). Examining Rh and Pt surfaces with and without adatoms and graphite surface with and without atomic vacancies, we confirmed that: i) bonds between undercoordinated atoms become shorter and stronger; ii) subjective polarization happens to the valence electrons of defects or adatoms by the densely entrapped bonding electrons, which in turn screens and splits the crystal field and hence the core band of the specimen.

[*] Ecqsun@ntu.edu.sg; ywang8@hnust.edu.cn

Bonds and electrons annexed the undercoordinated atoms dictate the unusual performance of materials at surface, defect, and at the nanoscale in many applications such as photoelectron emission,[1] local charge density,[2] mechanical strength and work function,[3] surface magnetism,[4] catalytic reactivity,[5-6] crystal growth,[7] adsorption,[8] oxidation,[9] decomposition,[10] doping,[11] interface formation,[12-13] wettability,[14-15] thermal stability,[16] etc. Although the chemistry and physics of materials associated with under-coordinated atoms have been extensively investigated, the laws governing the performance of such local bonds and electrons remain as yet to be established.[17-18] Collection and purification of such local, atomistic zone selective, and quantitative information having been increasingly demanded but it remains as yet the "dead corner" of the community.

As a powerful means, scanning tunneling micro/spectroscopy (STM/S) maps local electrons in the open side of a *too-thin* subatomic layer of a surface with energies of a few eVs cross Fermi energy ($E_F$). Understanding the intriguing STM/S attributes of high protrusions and the additional resonant spectral peak nearby $E_F$ at graphite atomic vacancy [19] and graphene edge,[20] for instances, as well as the driving force for the unusual protrusions and the $E_F$ resonance remains challenge. One urgently needs to identify what the "root" of the STM/S attributes is and what the quantitative information could be about the length and strength of the annexed bonds and the energies of the associated electrons. On the other hand, a photoelectron spectroscopy (PES such as ultraviolet or x-ray as the sources called UPS and XPS, respectively) probes statistic and volumetric information of electrons with binding energy in the valence band and below within a *too-thick* layer of 10 nanometers or thicker.[18, 21-23] In order to cope with this challenge, we have developed a special yet simple technique of zone-resolved photoelectron spectroscopy (ZPS), which has enabled us to realize the dreams.

We need to clarify first the reference point and the direction of the core level shift upon bulk formation and atomic undercoordination, as well as correlation between the bond length, bond energy and the core level shift as measured using XPS. According to the energy band theory of tight binding (TB) approach,[24] the energy shift of a specific vth core band from that of an isolated atom is proportional to the crystal potential energy at equilibrium or the cohesive energy per bond,

$$\Delta E_\nu(z) = E_\nu(z) - E_\nu(0) = \langle \nu,i | V_{cry}(r)(1+\Delta_H) | \nu,i+\nu,j \rangle \propto E_b(1+\Delta_H)[1+0(\langle \nu,i \| \nu,\rangle)]$$

(1)

Any perturbation to the crystal potential, $\Delta_H$, will shift the band away from the $E_\nu(0)$, assuming that the wave functions, for the core electrons at atomic sites i and i' ($\langle \nu,i|\nu,j\rangle=\delta_{ij}$) are strongly localized with insignificant overlapping.

Generally, an XPS spectrum can be decomposed into several Gaussian peeks representing contributions from atoms of different atomic coordination number, z. These components are therefore correlated by,

$$\frac{E_\nu(x)-E_\nu(0)}{E_\nu(12)-E_\nu(0)} = \frac{E_z}{E_b} = 1 + \Delta_H = \begin{cases} C_z^{-m} & (BOLS\ Entrapment) \\ \Delta_p + 1 & (Polarization) \end{cases}$$

(2)

The x represents z or p. $C_z = d_z/d_b = 2/\{1+\exp[(12-z)/(8z)]\}$ and $C_z^{-m} = E_z/E_b$ are, respectively, the coordination number (z)-dependent relative bond length and bond energy or the depth of the quantum entrapment, according to the bond order-length-strength correlation (BOLS).[25] The perturbation $\Delta_p = (E_\nu(p)-E_\nu(0))/(E_\nu(12)-E_\nu(0))-1$ arises from the effect of polarization on the core band with a component of $E_\nu(p)$. If the polarization-entrapment coupling effect is apparent, the term $C_z^{-m}$ is then replaced by $pC_z^{-m}$, the entrapped states will be moved up from the otherwise low-z entrapped energy. With the given energies of XPS components of z and z′ in decomposition, we can determine the referential $E_\nu(0)$ and the bulk shift, $E_\nu(12)-E_\nu(0)$,

$$\frac{E_\nu(z)-E_\nu(0)}{E_\nu(z')-E_\nu(0)} = \frac{C_z^{-m}}{C_{z'}^{-m}}, \text{ or, } E_\nu(0) = \frac{C_{z'}^m E_\nu(z') - C_z^m E_\nu(z)}{C_{z'}^m - C_z^m}$$

$$E_\nu(z) = E_\nu(0) + [E_\nu(12)-E_\nu(0)]C_z^{-m}$$

(3)

The BOLS-TB decomposition of the C 1s core level shifts of graphene edge, graphene, graphite, and diamond,[26] the 4f$_{7/2}$ level of Pt,[27] and the 3d$_{5/2}$ level of Rh[28] surfaces have derived the respective z-dependent core level energies,

$$E_\nu(z) = E_\nu(0) + [E_\nu(12)-E_\nu(0)]C_z^{-m} = \begin{cases} 282.57 + 1.32 C_z^{-2.56} & (C\ 1s) \\ 302.16 + 4.37 C_z^{-1} & (Rh\ 3d_{5/2}) \\ 67.21 + 3.28 C_z^{-1} & (Pt\ 4f_{7/2}) \end{cases} (eV)$$

(4)

**Figure 1** illustrate the correlation between the $E_v(0)$, the $E_v(12)$, and the z-dependent shift in comparison with those derivative from carbon allotropes.[25] These findings provide the foundations for the photoelectronic skinning of the adatoms, defects and the outermost atomic layers of a surface as demonstrated thus.

Instead of the tedious processing of XPS decomposition, we can purify the spectra due to atomic undercoordination using the ZPS method upon the spectral background correction and the peak area normalization. Figure 2 (upper part in each panel) shows the normalized XPS and the ZPS profiles of Pt(111)[29] and Rh(100)[30] surfaces with the indicated coverage of adatoms. The ZPS profiles were obtained by subtracting the normalized spectra of the surfaces with adatoms by the ones of clean surface. This processes derived the energy and effective z value for each component in both cases. The spectral valleys correspond to the bulk components and the peaks below the bulk valleys to the adatom induced quantum entrapment. The peak above the bulk component is due to the polarization of the adatoms. Instead of the entrapped states at z = 3, Rh adatoms show extra states at z = 4 ~ 6, with an addition of the polarized P states centered at 306.20 eV.

The difference between the Pt and the Rh spectra coincides exceedingly well with the BOLS expectation that only the otherwise conductive half-filled s-electron Rh($4d^8 5s^1$) can be polarized and anchored as monopoles to the adatoms. The possible reason for the loss of the initially trapped surface charge at z = 3 is that the electrons of adatoms are fully polarized, moving the otherwise entrapped states from z = 3 to z = 4 ~ 6 associated with the P states.

The raw XPS spectra, shown in Figure 3, were collected from (a) clean graphite (0001) surface at 25° and 75° emission angles (between the electronic beam and the surface normal) and from (b) the surface with and without $Ar^+$-spraying induced vacancy defects at 55°. One can hardly tell anything from the raw data but the ZPS brought a great difference. As an XPS collects more information from surface at large emission angles than those at small angles, the difference between the two spectra can discriminate the surface from the bulk, as shown in **Figure 3**(c). Likewise, the difference between the spectra collected before and after defect formation purifies the defect states as compared in **Figure 3**(c). The valleys centered around 284.20 eV and 284.40 eV correspond, respectively, to the removal graphite bulk and the mixture of surface-bulk; extra components are the energy states due to the quantum entrapment, $T_S(z \sim 3.1)$, of the outermost atomic layer and sites surrounding vacancy defects, $T_D(z \sim 2.2 \sim 2.4)$. G denotes the bulk graphite (z = 5.335). The P component at the upper band edge arises from the screening and splitting of the crystal potential by the Dirac-Fermi polarons, as STM/S identified,[19] that originate from the polarization of the dangling bond electrons by the densely entrapped ($T_D$ in the bottom of the core band) core electrons.[31]

We have thus demonstrated the power of the ZPS in gaining the local, atomistic zone selective, and quantitative information about the length and energy of the local bonds and the binding energy shift of electrons associated with the undercoordinated Pt and Rh adatoms, the outermost atomic layer of graphite surface with and without atomic vacancies. This effective yet simple means is shown to have solved the historical challenge and derive quantitative information at easy the energy shifts associated with atomic coordination and the effect of undercoordination induced quantum entrapment and the subjective polarization.

Financial supports from NSF (Nos. 11172254, 11002121and 10802071) of China and MOE (RG15/09), Singapore, and are gratefully acknowledged.
.

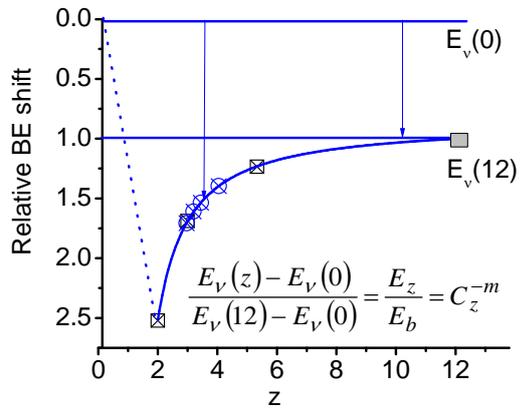

Figure 1 Illustration of the energy level of an isolated atom $E_v(0)$ and the relative shift induced by atomic undercoordination in comparison with experimental results [26, 32] of graphene edge (z=2), graphene (z=3), few-layer graphene, graphite and diamond (m = 2.56).

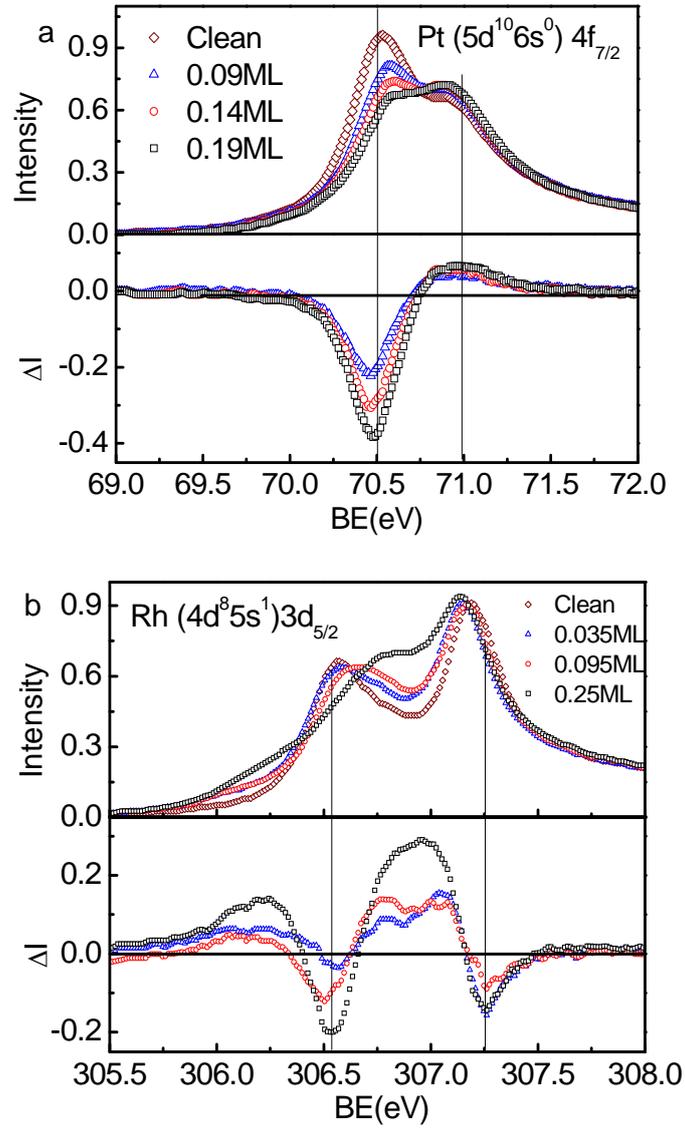

Figure 2 the normalized XPS spectra and the ZPS of (a) Pt(111) [29] and (b) Rh(100) [30] surface with indicated coverage of adatoms and the ZPS derivatives:

$$\left.\begin{array}{r}E_{Pt(111)4f_{7/2}}\\E_{Rh(100)3d_{5/2}}\end{array}\right\}(z)=\begin{cases}67.21+\begin{cases}3.28\equiv70.49 & (Bulk, z=12)\\3.70\equiv70.91 & (S1, z=4.25)\\3.87\equiv71.18 & (Adatoms, z=3.15)\\- & (Adatom\ Polaron)\end{cases}\\302.16+\begin{cases}4.37\equiv306.53 & (Bulk, z=12)\\4.99\equiv307.15 & (surface, z=4.0)\\5.35\equiv307.51 & (Adatoms, z=3.0)\\4.04=306.20 & (Adatom\ polaron)\end{cases}\end{cases}(eV)$$

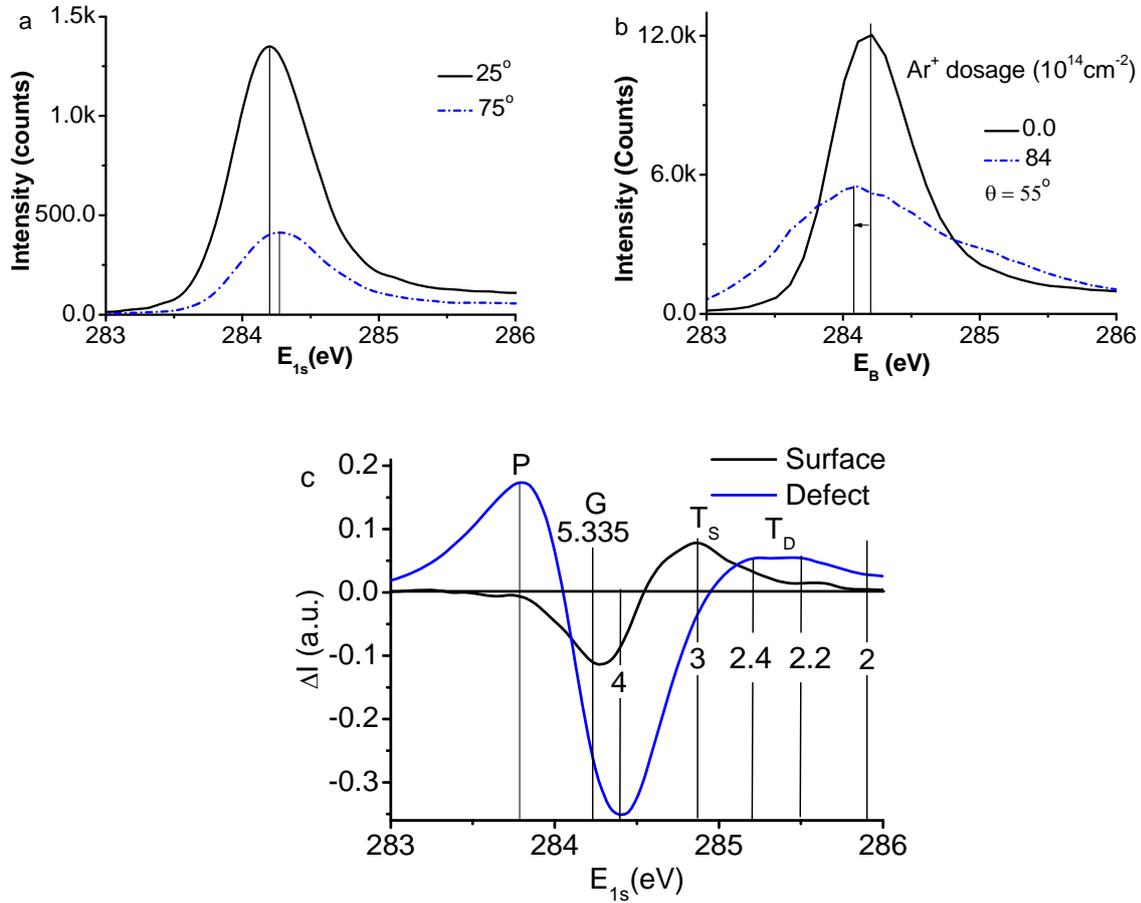

Figure 3 The raw XPS spectra collected from (a) defect-free graphite surface at 25° and 75° emission angles showing the positive shift and from (b) the defect surface at 55° of different Ar$^+$ ion doses showing the broadening and negative shift. (c) The purified C 1s ZPS from the surface with (9×10$^{14}$ cm$^{-2}$ dosed Ar$^+$ ion) and without defects clarifies the surface and defect states and the quantitative information of:

$$E_{C(0001)_{1s}}(z) = 282.57 + \begin{cases} 1.32 \equiv 284.20 & (Bulk, z = 5.335) \\ 2.12 \equiv 284.75 & (S1, z = 3.1) \\ 2.84 \equiv 285.41 & (Adatoms, z = 2.3) \\ 1.23 \equiv 283.80 & (Vacancy\ polaron) \end{cases} (eV)$$